\documentclass{ws-ijmpd}
\usepackage[super,compress]{cite}
\begin{document}

\markboth{A. Paliathanasis, S. Basilakos, M. Tsamparlis}
{Comment on [Int. J. Mod. Phys. D 25 (2016) 1650051]}

%
\catchline{}{}{}{}{}
%

\title{Comment on \textquotedblleft A study of phantom scalar field cosmology using
Lie and Noether symmetries\textquotedblright\ [Int. J. Mod. Phys. D 25 (2016) 1650051]}

\author{ANDRONIKOS PALIATHANASIS}

\address{Instituto de Ciencias F\'{\i}sicas y Matem\'{a}ticas, \\ Universidad Austral de
Chile, Valdivia, Chile\\
anpaliath@phys.uoa.gr}

\author{SPYROS BASILAKOS}

\address{Academy of Athens, Research Center for Astronomy and Applied Mathematics, \\
Soranou Efesiou 4, 11527, Athens, Greece\\
svasil@academyofathens.gr}

\author{MICHAEL TSAMPARLIS}

\address{Faculty of Physics, Department of Astrophysics - Astronomy - Mechanics, \\
University of Athens, \\ Panepistemiopolis, Athens 157 83, Greece \\
mtsampa@phys.uoa.gr}

\maketitle

\begin{history}
\received{Day Month Year}
\revised{Day Month Year}
\end{history}

\begin{abstract}
We show that the recent results of \ [Int. J. Mod. Phys. D 25 (2016) 1650051]
on the application of Lie/Noether symmetries in scalar field cosmology are
well-known in the literature while the problem could have been solved easily
under a coordinate transformation. That follows from the property, that the
admitted group of invariant transformations of dynamical system is independent
on the coordinate system.

\end{abstract}


\ccode{PACS numbers:98.80.-k, 95.35.+d, 95.36.+x}


In \cite{dutta} the authors consider the action%
\begin{equation}
S=\int d^{4}x\sqrt{-g}\left[  R-\frac{1}{2}\lambda\left(  \phi\right)
g^{\mu\nu}\phi_{;\mu}\phi_{;\nu}-V\left(  \phi\right)  \right]  +S_{m}%
\label{eq.1}%
\end{equation}
where $S_{m}$, corresponds to the matter source, and $\lambda\left(
\phi\right)  $ is an unknown function in a spatially flat
Friedmann--Robertson--Walker (FRW) spacetime with signature $\left(
-,+,+,+\right)  $, scale factor $a\left(  t\right)  $, and a perfect fluid
with constant equation of state parameter, $p=\left(  \gamma-1\right)  \rho$.
From this action for the lapse time $N\left(  t\right)  =1$, and for the
comoving observers,~$u^{a}=\delta_{t}^{a}$, it follows that the Lagrangian of
the field equations is%
\begin{equation}
L\left(  a,\dot{a},\phi,\dot{\phi}\right)  =-3a\dot{a}^{2}+\frac{1}{2}%
\lambda\left(  \phi\right)  a^{3}\dot{\phi}^{2}-a^{3}V\left(  \phi\right)
-\rho_{m0}a^{3\left(  1-\gamma\right)  },\label{eq.2}%
\end{equation}
while the corresponding field equations are the \textquotedblleft
Hamiltonian\textquotedblright\ function, which is the first Friedmann's
equation
\begin{equation}
3a\dot{a}^{2}-\frac{1}{2}\lambda\left(  \phi\right)  a^{3}\dot{\phi}^{2}%
-a^{3}V\left(  \phi\right)  -\rho_{m0}a^{3\left(  1-\gamma\right)
}=0,\label{eq.3}%
\end{equation}
and the Euler-Lagrange equations of Lagrangian (\ref{eq.2}) for the variables
$\left\{  a,\phi\right\}  $. In \cite{dutta}, the authors claim that the
existence of Lie point symmetries for the field equations or Noether point
symmetries for the Lagrangian (\ref{eq.2}) result in constraints which provide
the unknown parameters of the model, that is $\gamma$, $V\left(  \phi\right)
$ and the function $\lambda\left(  \phi\right)  $. The purpose of this short
note is to show that this is not true and the correct answer is that the only
parameters which can be constrained are $\gamma$, and $V\left(  \phi\right)
$, while the last will be $V\left(  \lambda\left(  \phi\right)  \right)  $, or
more precisely, $V\left(  \phi\right)  =V\left(  \int\sqrt{\lambda\left(
\phi\right)  }d\phi\right)  $.

First we note that the Lie algebra of the Lie/Noether symmetries of a
dynamical system are independent on the coordinate system. Therefore if in the
Lagrangian (\ref{eq.2}) we define the new field $\psi,~$such as $d\psi
=\sqrt{\lambda\left(  \phi\right)  }d\phi$ we obtain%
\begin{equation}
L\left(  a,\dot{a},\psi,\dot{\psi}\right)  =-3a\dot{a}^{2}+\frac{1}{2}%
a^{3}\dot{\psi}^{2}-a^{3}V\left(  \psi\right)  -\rho_{m0}a^{3\left(
1-\gamma\right)  }\label{eq.10}%
\end{equation}
where $V\left(  \psi\right)  =V\left(  \int\sqrt{\lambda\left(  \phi\right)
}d\phi\right)  .$ This is the classical Lagrangian of a minimally coupled
scalar field cosmological model in a spatially flat FRW spacetime. Furthermore
in the case in which $\psi\rightarrow i\psi$,~or $\lambda\left(  \phi\right)
<0,$ we have the Lagrangian of a phantom field.

Hence the symmetry analysis of (\ref{eq.2}) or (\ref{eq.10}) will give the
same results, however of a different form of the potentials $V\left(
\psi\right)  $, $V\left(  \phi\right)  $, but which they will be related under
the transformation $\phi\rightarrow\psi$. Furthermore, if the latter
transformation is not complex, the solution for the scale factor will be
exactly the same.

Concerning the application of Noether symmetries of the gravitational
Lagrangian (\ref{eq.2}), the potential $V(\phi)$ in (\ref{eq.1}) the authors
of \cite{dutta} find is the Unified Dark Matter potential (UDM) (for instance
see \cite{berta,gorini,basill}). For instance, for $\lambda\left(
\phi\right)  =\frac{\lambda_{0}}{\phi^{2}},~$in (\ref{eq.2}), they have found
the potentials
\begin{align}
V\left(  \phi\right)   &  =V_{0}\sinh^{2}\left(  \frac{3}{8}\left\vert
\lambda_{0}\right\vert \ln\phi+p_{1}\right)  ,~\lambda_{0}>0\\
V\left(  \phi\right)   &  =V_{0}\sin^{2}\left(  \frac{3}{8}\left\vert
\lambda_{0}\right\vert \ln\phi+p_{1}\right)  ,~\lambda_{0}<0
\end{align}
which under the coordinate transformation $\psi=\ln\phi$, which is the same
result when $\lambda\left(  \phi\right)  =\lambda_{0}$, i.e. a constant,
becomes
\begin{align}
V\left(  \psi\right)   &  =V_{0}\sinh^{2}\left(  \frac{3}{8}\psi+p_{1}\right)
\\
V\left(  \psi\right)   &  =V_{0}\sin^{2}\left(  \frac{3}{8}\psi+p_{1}\right)
.
\end{align}

We would like to remark that the application of the complete Noether's
theorem in scalar field cosmology can be found in \cite{paper1} and the
application of Lie point symmetries in \cite{paper2}, hence the results of
\cite{dutta} are not new in the literature.

Furthermore we wish to draw attention to the paper by Capozziello et al.
\cite{cap96} on Scalar-tensor theories in which an extended discussion on
the application of Noether symmetries in cosmology is presented; which
includes also the results on Noether symmetries of \cite{dutta}.

Finally we refer the authors of \cite{dutta} to the original work on symmetries of
differential equations of S. Lie \cite{lie} and its application to the
Action Integral which has been done by E. Noether \cite{noe}.
{\bf {Acknowledgments:}}
AP acknowledges financial support of FONDECYT postdoctoral grant no. 3160121.

\end{document}